\documentclass[pra,aps,twocolumn,showpacs]{revtex4}

\usepackage{amssymb}

\usepackage{graphicx}
\usepackage{bm}

\begin{document}

\title{Mixed state localizable entanglement for continuous variables}

\author{Ladislav Mi\v{s}ta, Jr.}
\affiliation{Department of Optics, Palack\'{y} University,
17. listopadu 50, 77200 Olomouc, Czech Republic}

\author{Jarom\'{\i}r Fiur\'{a}\v{s}ek}
\affiliation{Department of Optics, Palack\'{y} University,
17. listopadu 50, 77200 Olomouc, Czech Republic}

\begin{abstract}
We investigate localization of entanglement of multipartite mixed  Gaussian states
into a pair of modes by local Gaussian measurements on the remaining modes and classical communication. We provide a detailed proof that for arbitrary symmetric Gaussian state
maximum entanglement can be localized by homodyne detection of either amplitude or phase quadrature on each mode. We then consider arbitrary mixed three-mode Gaussian states and show that the optimal Gaussian measurement on one mode yielding maximum entanglement among the other two modes can be determined by calculating roots of a high-order polynomial. Finally, we discuss localization of entanglement with single-photon detection.
\end{abstract}


\maketitle


\section{Introduction}

Continuous quantum variables have emerged during recent years as a very promising platform
for quantum communication and information processing \cite{Braunstein05,Cerf07}. Using coherent and squeezed states of light instead of single photons for encoding and processing quantum information possesses distinct advantages. The entangled squeezed states can be generated deterministically in optical parametric amplifiers and can be detected with very high efficiency using balanced homodyne detectors. This should be contrasted with probabilistic generation of entangled photon pairs in the process of spontaneous parametric downconversion and comparatively low efficiency of the single-photon detectors. The favourable scaling of continuous variable systems greatly facilitates preparation of complex multimode
entangled Gaussian squeezed states. Various three- and four-mode entangled squeezed states were demonstrated experimentally \cite{Aoki03,Su07,Yukawa08,Tan08}
and exploited in elementary quantum teleportation networks \cite{Yonezawa04,Yonezawa07}.  Properties of specific multimode entangled Gaussian states \cite{Wolf04} such as symmetric states \cite{Adesso04,Adesso05,Serafini05} or cluster states \cite{Zhang06,Menicucci06,vanLoock07a,vanLoock07b} were theoretically analyzed in detail.

The multimode entangled states could form a backbone of an advanced quantum communication network where each node contains a part of the entangled system. An interesting and important question is how much bipartite entanglement between two nodes of the network can be \emph{localized} on average by performing local measurements in each node and exchanging the measurement results via classical communication. This so-called localizable entanglement \cite{Verstraete04} has been originally introduced and studied for discrete variable systems, in particular quantum spin chains. Recently, we have generalized the notion of localizable entanglement to continuous-variable systems \cite{Fiurasek07} and investigated the localization of entanglement of multimode Gaussian states via local Gaussian measurements. We have proved that for pure states it is always optimal to perform local homodyne detections, i.e. projections onto infinitely squeezed states.
Moreover, we have also shown that non-Gaussian measurements such as photon counting allow us to localize more entanglement in certain cases.

In this paper we investigate localization of entanglement for mixed multimode Gaussian states. We consider entanglement localization by Gaussian measurements that preserve the Gaussian form of the resulting two-mode bi-partite state. We quantify the entanglement by logarithmic negativity which is a computable entanglement  measure for Gaussian states \cite{Vidal02}. We first present a detailed proof that homodyne detection is optimal for localization of entanglement of mixed symmetric Gaussian states \cite{Adesso04,Adesso05,Serafini05,Adesso06}.
We then consider arbitrary three-mode mixed states and, somewhat surprisingly, find that in this case
maximum entanglement among two modes can be sometimes localized by eight-port homodyne detection, i.e. projection onto coherent states, on the third mode. We thus demonstrate that the homodyne detection is not always an optimal strategy if we deal with general mixed Gaussian states. Finally, we go beyond the realm of Gaussian measurements and analyze localization of entanglement with realistic single photon detectors that can only distinguish between presence or absence of photons in the light beam but cannot resolve the number of photons in the beam. We show that in principle even such low-efficiency realistic single-photon detectors may be sometimes advantageous for entanglement localization compared to Gaussian measurements.

The paper is structured as follows. In Sec. II we provide a brief summary of the main properties of Gaussian states. Then in Sec. III we define the Gaussian localizable entanglement and discuss some of its features. Section IV contains a detailed analytical proof of optimality of local homodyne detection for localization of entanglement of multimode generally mixed symmetric Gaussian states. In Sec. V we study the localization of entanglement for generic three-mode mixed Gaussian states. Localization of entanglement by single-photon detection is addressed in Sec. VI. Finally, the conclusions are drawn in Sec. VII.


\section{Gaussian states and measurements}

We deal with quantum systems with infinite-dimensional Hilbert space such as
modes of the electromagnetic field. An $N$-mode system can be described by the quadrature
operators $x_{j},p_{j}$, $j=1,\ldots,N$ satisfying the canonical commutation relations
$\left[x_{j},p_{k}\right]=i\delta_{jk}$. Arranging the operators into the vector
$R=\left(x_{1},p_{1},\ldots,x_{N},p_{N}\right)^{T}$ the commutation relations
can be compactly expressed as $\left[R_{j},R_{k}\right]=i\Omega_{jk}$, where
$\Omega=\oplus_{j=1}^{N}J$ and
\begin{equation}\label{J}
J= \left(
\begin{array}{cc}
0  & 1 \\
-1 & 0
\end{array}
\right).
\end{equation}
States of the $N$-mode system can be represented in the $2N$-dimensional real phase space by the Wigner
function \cite{Wigner_32}. Among all possible states, the states with Gaussian-shaped Wigner function are particularly important. These so-called Gaussian states can be easily prepared experimentally using  squeezers and passive linear optics and can be efficiently handled theoretically.
Each Gaussian state $\rho$ is fully characterized by the first moments of the quadrature operators $\langle{R}_{i}\rangle=\mbox{Tr}(\rho R_{i})$,
$i=1,\ldots,2N$ and by the covariance matrix (CM) $\gamma$ with elements $\gamma_{ij}=\langle\left\{\Delta R_{i},\Delta R_{j}\right\}\rangle$,
where $\Delta R_{i}=R_{i}-\langle R_{i}\rangle$ and $\left\{A,B\right\}=AB+BA$. For any CM $\gamma$ there is a symplectic matrix $S$, i.e. a $2N\times2N$ real matrix satisfying the condition $S\Omega S^{T}=\Omega$, that brings the CM to the normal form \cite{Williamson_36} $S\gamma S^{T}=\mbox{diag}(\nu_{1},\nu_{1},\ldots,\nu_{N},\nu_{N})$. The quantities  $\nu_{i}\geq1$, $i=1,\ldots,N$ are the so called symplectic
eigenvalues and can be calculated from the eigenvalues of the matrix $\Omega\gamma$ that read as
$\left\{\pm i\nu_{1},\ldots,\pm i\nu_{N}\right\}$ \cite{Vidal02}.

Gaussian states of two or more modes can be entangled. In the case of just two modes $A$ and $B$ the
state $\rho_{AB}$ is entangled if and only if its partial transposition with respect to mode $B$ $\rho_{AB}^{T_B}$ is
not a positive-semidefinite matrix \cite{Peres_96,Horodecki_96,Simon_00,Duan_00}. On the level of CMs the partial
transposition operation is represented by the diagonal matrix $\Lambda=\mbox{diag}\left(1,1,1,-1\right)$ \cite{Simon_00} and the
CM $\gamma_{AB}$ of the state $\rho_{AB}$ is transformed by matrix congruence $\tilde{\gamma}_{AB}=\Lambda\gamma_{AB}\Lambda^T$.
Let $\mu_{1},\mu_{2}$ denote the  symplectic eigenvalues of $\tilde{\gamma}_{AB}$ and we assume without loss of any generality that $\mu_2 \leq \mu_1$.
The Gaussian state $\rho_{AB}$ is entangled if and only if $\tilde{\gamma}_{AB}$ violates the generalized Heisenberg uncertainty relation $\gamma+i\Omega \geq 0$ \cite{Simon_00} or equivalently if and only if $\mu_{2}<1$. It is convenient to express the CM $\gamma_{AB}$ in a
block form with respect to $A|B$ splitting,
\begin{equation}
\gamma_{AB}= \left(
\begin{array}{cc}
A  & C \\
C^{T} & B
\end{array}
\right).
\end{equation}
The symplectic eigenvalues of the matrix $\tilde{\gamma}_{AB}$ can be calculated using the formula \cite{Vidal02}
\begin{equation}\label{symplectic}
\mu_{1,2}=\sqrt{\frac{\delta\pm\sqrt{\delta^{2}-4\Delta}}{2}},
\end{equation}
where $\delta=\mbox{det}A+\mbox{det}B-2\mbox{det}C$ and $\Delta=\mbox{det}\gamma_{AB}$ are the
so-called symplectic invariants.
 The entanglement of a bipartite state $\rho_{AB}$ can be quantified by
the logarithmic negativity $E_{\mathcal{N}}[\rho_{AB}]=\log_2||\rho_{AB}^{T_B}||_1$.
The logarithmic negativity has the important advantage that it can be computed analytically for an arbitrary mixed two-mode Gaussian state \cite{Vidal02}  and we have
\begin{equation}
E_{\mathcal{N}}[\rho_{AB}]=\max(0,-\log_2 \mu_2),
\label{EN}
\end{equation}
which is a monotonously decreasing function of the lower symplectic
eigenvalue $\mu_{2}$ of the matrix $\tilde{\gamma}_{AB}$ \cite{Adesso06}.

Arbitrary quantum measurement can be described by positive operator valued measure  (POVM) whose elements $\Pi_j$ are positive semidefinite operators satisfying the completeness condition $\sum_j \Pi_j =\openone$, where $\openone$ denotes the identity operator. Gaussian measurement on some mode $C$ is any measurement whose POVM elements $\Pi_j$ are represented by Gaussian Wigner functions. This means that any operator $\Pi_j$ is proportional to a density matrix of some (generally mixed) Gaussian state.
In practice, any Gaussian measurement on optical modes can be performed using
 auxiliary modes prepared in vacuum states, passive and active linear optics (beam splitters, phase shifters and squeezers) and balanced homodyne detection on each mode.
The corresponding continuous Gaussian POVM characterizing the measurement on $C$ reads
\begin{equation}
\Pi_C(\alpha)= \frac{1}{\pi} D_C(\alpha)\Pi_C^0 D_C^\dagger(\alpha).
\label{Pialpha}
\end{equation}
Here $D_C(\alpha)=\exp(\alpha a_C^\dagger-\alpha^\ast a_C)$ denotes the displacement
operator and $a_C=\frac{1}{\sqrt{2}}(x_C+ip_C)$ is the annihilation operator.
$\Pi_C^0$ is a density matrix of a Gaussian state with covariance matrix
$\gamma_{C}^M$ and zero mean values of the quadratures, and the displacement $\alpha$ is determined by the outcomes of the homodyne detectors. Note that $\gamma_{C}^M$ is fixed by the structure of the measurement setup and does not depend
on the measurement outcomes. The displacement operation is an irreducible representation of the Weyl-Heisenberg
group and Schur's lemma together with the normalization $\mathrm{Tr}[\Pi_C^0]=1$ implies that
\begin{equation}
\int \Pi_C(\alpha) d^2\alpha= \openone.
\label{completness}
\end{equation}
This ensures the completness of the POVM (\ref{Pialpha}).


\section{Gaussian localizable entanglement}

We consider $N$-mode Gaussian state $\rho_{AB\bm{C}}$ shared among $N$ parties
$A$, $B$, $C_j$, $j=1,\ldots, N-2$, with each party possessing a single mode.
Parties $C_j$ attempt to increase the entanglement between $A$ and $B$ by performing
\emph{local} Gaussian measurements and communicating the measurement outcomes to
$A$ and $B$. We are interested in the maximum amount of entanglement that can be localized
on average in the modes $A$ and $B$ in this way.
We will use the easily computable logarithmic negativity to quantify the entanglement of mixed Gaussian states.

The elements of the total POVM describing measurement on all modes $C_j$
can be written as product of the single-site elements,
\[
\Pi_{\bm{C}}(\bm{\alpha})=\bigotimes_{j=1}^{N-2} \Pi_{C_j}(\alpha_j),
\]
where $\bm{\alpha}=(\alpha_1,\cdots,\alpha_{N-2})$. The density matrix of the Gaussian state of modes $A$ and $B$ conditional on the particular measurement outcome  $\bm{\alpha}$ can be expressed as
\[
\sigma_{AB}(\bm{\alpha})=\frac{\mathrm{Tr}_{\bm{C}}[\openone_{AB}\otimes \Pi_{\bm{C}}(\bm{\alpha})\,\rho_{AB \bm{C}}]}{\mathrm{Tr}[\openone_{AB}\otimes \Pi_{\bm{C}}(\bm{\alpha})\,\rho_{AB \bm{C}}]},
\]
where $\mathrm{Tr}_{\bm{C}}$ denotes partial trace over the modes $C_j$.
The entanglement of the Gaussian state $\sigma_{AB}$ depends only on its covariance matrix and not on its coherent displacement, because mean values of all quadratures can be set to zero by means of local unitary displacement operations that do not change the entanglement. The covariance matrix of $\sigma_{AB}(\bm{\alpha})$
depends only on the covariance matrix of the initial $N$-mode state $\gamma_{AB\bm{C}}$ and on the covariance matrices
$\gamma_{C_j}^M$ characterizing the Gaussian measurements on each mode $C_j$. This means that entanglement of $\sigma_{AB}(\bm{\alpha})$ does not depend on $\bm{\alpha}$ which determines only the coherent displacement of the conditionally prepared state.
We have that $E_{\mathcal{N}}[\sigma_{AB}(\bm{\alpha})]=E_{\mathcal{N}}[\sigma_{AB}(\bm{0})]$, $\forall \bm{\alpha}$.
Moreover, any projection onto mixed Gaussian state can be expressed as a Gaussian mixture of projections onto pure Gaussian states. Clearly, such incoherent mixing can only reduce the localizable entanglement.
It follows that in order to determine the maximum Gaussian localizable
entanglement among modes $A$ and $B$, $E_{L,G}[\rho_{AB\bm{C}}]$, it suffices to optimize the logarithmic negativity $E_{\mathcal{N}}[\sigma_{AB}(\bm{0})]$ over all local projections onto pure single-mode squeezed vacuum states \cite{Fiurasek07},
\begin{equation}
E_{L,G}[\rho_{AB\bm{C}}]=\max_{\gamma_{C_j}^M}\left\{E_{\mathcal{N}}[\sigma_{AB}(\bm{0})]\right\},
\label{ELG}
\end{equation}
where $\gamma_{C_j}^M+iJ \geq 0$ and $\det(\gamma_{C_j}^M)=1$.
The definition of logarithmic negativity (\ref{EN}) implies that the
maximization (\ref{ELG}) amounts to minimization of the smaller symplectic eigenvalue $\mu_2$ of the covariance matrix of the partially transposed state $\sigma_{AB}^{T_B}(\bm{0})$ over
all local projections of $C_j$ onto pure Gaussian squeezed vacuum states.

As a side remark, we note that the local projections onto pure Gaussian states
are optimal among a wider class of POVMs whose elements can be written as a product of
single-site operators, $\Pi_{\bm{C}}(\bm{z})=\otimes_{j=1}^{N-2}\Pi_{C_j}(z_j)$. Here $\bm{z}$ is
a multi-index labeling the POVM elements and the completness of the POVM dictates
that $\int_{\bm{z}}\Pi_{\bm{C}}(\bm{z}) d \bm{z}=\openone$. Each element $\Pi_{C_j}(z_j)$
is Gaussian but the overall structure of the POVM can be otherwise arbitrary.
We denote by $\sigma_{AB}(\bm{z})$ the normalized conditionally prepared state
of $A$ and $B$. It immediately follows that
\begin{equation}
E_{L,G} \leq \max_{\otimes_{j=1}^{N-2}
\Pi_{j}(z_j)}E_{\mathcal{N}}[\sigma_{AB}(\bm{z})]=E_{\mathcal{N}}[\sigma_{AB}(\bm{z}^0)].
\label{ELGbound}
\end{equation}
where the  maximization has to be carried out over \emph{all}
Gaussian POVM elements $\Pi_{j}(z_j)$ and $\bm{z}^0$  labels the POVM element
which maximizes  the entanglement between $A$ and $B$.
Using  $\Pi_{j}(z_{j}^0)$ as a seed element for the construction of local Gaussian
POVM (\ref{Pialpha}) on mode $C_j$  we construct a local Gaussian  measurement which
achieves the localizable entanglement $E_{\mathcal{N}}[\sigma_{AB}(\bm{z}^0)]$ and thereby
maximizes $E_{L,G}$.

Recently, the problem of localizable entanglement was studied for $N$-mode pure Gaussian states
for which it was shown that maximum entanglement is localized by homodyne detection \cite{Fiurasek07}.
There it was also noted that homodyne detection is optimal in the case of $N$-mode
mixed fully symmetric Gaussian states. A detailed analytical proof of the latter statement as well as
its generalization to $N$-mode bisymmetric states is given in the following section.

\section{Symmetric N-mode states}\label{symmetric}

The fully symmetric states are invariant under the exchange of any pair of modes \cite{Adesso04,Adesso05}.
Owing to the symmetry these states have the highly symmetric CM of the form:
\begin{equation}
\gamma_{\mathrm{sym}}=\left(
\begin{array}{cccc}
\beta & \epsilon & \ldots & \epsilon  \\
\epsilon & \beta &  \epsilon & \vdots  \\
\vdots & \epsilon & \ddots & \epsilon \\
\epsilon & \cdots & \epsilon & \beta
\end{array}
\right),
\label{gammasym}
\end{equation}
where $\beta$ and $\epsilon$ denote symmetric $2\times 2$ matrices. The matrices $\beta$ and $\epsilon$ can
be diagonalized simultaneously by local canonical transformations that do not affect the entanglement so
that without loss of any generality we may assume that $\beta=\mbox{diag}(b,b)$ and
$\epsilon=\mbox{diag}(\epsilon_{1},\epsilon_{2})$ \cite{Serafini05}.

We want to find the minimum of the lower symplectic eigenvalue $\mu_2$ of the matrix $\tilde{\gamma}_{AB}$
associated with state of modes $A$ and $B$ after local
Gaussian measurements on remaining $N-2$ modes $C_j$. The minimization needs to be carried out over all such measurements.
This optimization task can be greatly simplified if we notice that the state
(\ref{gammasym}) can be brought by 
unitary transformations
$U_{AB}$ and $U_{\bf {C}}$ on modes $AB$ and $C_{1}C_{2}\ldots
C_{N-2}$, respectively, into the block diagonal form with $N-2$
blocks $2\times2$
$\gamma_{B}=\gamma_{C_{2}}=\ldots=\gamma_{C_{N-2}}=\beta-\epsilon$
and one $4\times 4$ block
\begin{eqnarray}
\gamma_{AC_1}=\left(
\begin{array}{cc}
\beta+\epsilon & \sqrt{2(N-2)} \epsilon \\
\sqrt{2(N-2)} \epsilon & \beta +(N-3)\epsilon
\end{array}
\right). \label{gammaAC}
\end{eqnarray}
The transformation $U_{\bf{C}}$ can be physically realized by the
following sequence of $N-3$ beam splitters
\begin{eqnarray}\label{BSarray}
U_{\bf{C}}&\equiv&
B_{C_1C_2}\left(\sin^{-1}\frac{1}{\sqrt{N-2}}\right)B_{C_1C_3}\left(\sin^{-1}\frac{1}{\sqrt{N-3}}\right)\nonumber\\
&&\times\ldots\times
B_{C_1C_{N-2}}\left(\sin^{-1}\frac{1}{\sqrt{2}}\right),
\end{eqnarray}
while the transformation $U_{AB}$ can be performed by a single
balanced beam splitter, i.e. $U_{AB}=B_{AB}(\pi/4)$. Here
$B_{ij}(\theta)$ denotes a standard beam splitter transformation
between modes $i$ and $j$
\begin{eqnarray}\label{beamsplitter}
a_{i}&\rightarrow& a_{i}\cos\theta+a_{j}\sin\theta\nonumber\\
a_{j}&\rightarrow& a_{i}\sin\theta-a_{j}\cos\theta,
\end{eqnarray}
where $a_k$ ($a_{k}^{\dag}$), $k=i,j$ denote standard annihilation
(creation) operators.

The interference on the array of beam splitters transforms the $N$-mode symmetric state
into a product of $N-2$ uncorrelated single-mode states of modes $B$ and $C_j$, $j>1$, and a (possibly entangled)  two-mode state of $A$ and $C_1$.
In this way our task boils down to finding
 a Gaussian measurement on a single mode $C_1$ that will prepare
the mode $A$ in a state that will give after mixing
with a state with covariance matrix $\gamma_{B}=\beta-\epsilon$ on a balanced beam
splitter $U_{AB}$ the maximum entanglement. Projection of mode
$C_{1}$ onto a Gaussian state with CM $\gamma_{C}^{M}$ prepares mode
$A$ in the state with CM \cite{Giedke02,Eisert02}
\begin{equation}\label{gammaAprimed}
\gamma_{A}'=\beta+\epsilon-2(N-2)\epsilon[\beta+(N-3)\epsilon+\gamma_{C}^{M}]^{-1}\epsilon.
\end{equation}
Without loss of any generality we can assume $\gamma_{C}^{M}$ to be CM of a pure squeezed state,
\begin{equation}\label{tildegammaC}
\gamma_{C}^{M}=U(\theta) V(r) U^T(\theta),
\end{equation}
where
\begin{equation}
U(\theta)= \left(
\begin{array}{cc}
\cos\theta  & \sin \theta \\
-\sin\theta & \cos\theta
\end{array}
\right),
\quad
V(r)= \left(
\begin{array}{cc}
e^{2r}  & 0 \\
0 & e^{-2r}
\end{array}
\right).
\label{UVdefinition}
\end{equation}
Mixing of the states with CMs $\gamma_{A}'$ and $\gamma_{B}$ on a balanced beam splitter yields
a fully symmetric state with CM $\kappa_{AB}$ with the diagonal and off-diagonal blocks of the form
$\omega=(\gamma_{A}'+\gamma_{B})/2$ and $\zeta=(\gamma_{A}'-\gamma_{B})/2$, respectively. The
symplectic eigenvalues $\mu_{1,2}$ of the matrix $\tilde{\kappa}_{AB}$ corresponding to the
partially transposed state can be calculated with the help of the formula (\ref{symplectic})
where $\delta=2(\mbox{det}\omega-\mbox{det}\zeta)$ and $\Delta=\mbox{det}\kappa_{AB}$.
Making use of the formula for determinant of a sum of two $2\times 2$ matrices $P$ and $Q$,
\begin{equation}
\det(P+Q)=\det{P}+\det{Q}+\mathrm{Tr}[PJQ^{T}J^{T}],
\label{detsum}
\end{equation}
where $J$ is given in Eq.~(\ref{J}),
one finally finds the square of the minimum symplectic eigenvalue $\mu_{2}$ to be
\begin{equation}\label{mineig}
\mu_2^{2}=\mbox{det}\left(\beta-\epsilon\right)\lambda_{\rm min},
\end{equation}
where
\begin{equation}\label{lambda}
\lambda_{\rm
min}=\min_{\gamma_{C}^{M}}\left\{\mbox{eig}\left[{\left(\beta-\epsilon\right)}^{-\frac{1}{2}}\gamma_{A}'{\left(\beta-\epsilon\right)}^{-\frac{1}{2}}\right]\right\},
\end{equation}
and $\mbox{eig}(\cal{A})$ stands for the set of eigenvalues of
the matrix $\cal{A}$. Now we seek $\lambda_{\rm min}$ such that
for any admissible  $\gamma_{A}'$ it holds
\begin{equation}\label{inequality1}
{\left(\beta-\epsilon\right)}^{-\frac{1}{2}}\gamma_{A}'{\left(\beta-\epsilon\right)}^{-\frac{1}{2}}-\lambda_{\rm
min}\openone\geq0,
\end{equation}
and there is $\gamma_{A}'$ for which the matrix on the left-hand
side of the inequality has some eigenvalue equal to zero.
Multiplying the inequality (\ref{inequality1}) from the left and right by the positive definite matrix $(\beta-\epsilon)^{\frac{1}{2}}$, substituting for $\gamma_{A}'$
from Eq.~(\ref{gammaAprimed}) and multiplying it again from the
left and right by the matrix $\epsilon^{-1}$ one gets the
equivalent inequality
\begin{equation}\label{inequality2}
X-Y^{-1}\geq0,
\end{equation}
where
\begin{eqnarray}\label{XY}
X&=&\frac{\epsilon^{-1}\left[\beta+\epsilon-\lambda_{\rm
min}(\beta-\epsilon)\right]\epsilon^{-1}}{2(N-2)},\nonumber\\
Y&=&\beta+(N-3)\epsilon+\gamma_{C}^{M}.
\end{eqnarray}
Multiplying now inequality (\ref{inequality2}) from both sides by
$X^{-\frac{1}{2}}$ then by
$(X^{\frac{1}{2}}YX^{\frac{1}{2}})^{\frac{1}{2}}$ and finally
again by $X^{-\frac{1}{2}}$ one arrives at the equivalent
inequality
\begin{equation}\label{inequality3}
Y-X^{-1}\geq0.
\end{equation}
Substituting here from Eq.~(\ref{XY}), using
the decomposition (\ref{tildegammaC}) for $\gamma_{C}^{M}$  and taking into account that
$\beta$ and $\epsilon$ are diagonal, one can rewrite the inequality (\ref{inequality3})
in the following simpler form,
\begin{equation}\label{inequality4}
Z\equiv\mbox{diag}(d_1,d_2)+U(\theta)V(r)U^{T}(\theta)\geq0.
\end{equation}
Here  $\mbox{diag}(d_1,d_2)$ is a diagonal matrix with entries
\begin{equation}\label{dj}
d_{j}=b+(N-3)\epsilon_{j}-\frac{2(N-2)\epsilon_{j}^{2}}{b+\epsilon_{j}-\lambda_{\rm
min}(b-\epsilon_{j})},\quad j=1,2.
\end{equation}
The eigenvalues of the matrix $Z$ are lower bounded as
\begin{equation}\label{inequality5}
\mbox{min}[\mbox{eig}(Z)]\geq \mbox{min}(d_1,d_2)+e^{-2r},
\end{equation}
and the bound is saturated by choosing $\theta=0$ if $d_{1}\geq d_{2}$ and $\theta=\pi/2$
if $d_{1}<d_{2}$. The expression on the right-hand side of Eq. (\ref{inequality5}) is further lower bounded by $\mbox{min}(d_{1},d_{2})$
and this ultimate lower bound is achieved in the limit $r\rightarrow\infty$ which reveals that the optimal measurement
on $C_1$ localizing maximum entanglement between modes $A$ and $B$ is the homodyne detection of quadrature $p$
provided that $d_{1}\geq d_{2}$ and of quadrature $x$ provided that $d_{1}<d_{2}$. The sought quantity
$\lambda_{\rm min}$ can be derived from the condition $\mbox{min}(d_{1},d_{2})=0$ which gives using
Eq.~(\ref{dj}) after some algebra
\begin{equation}\label{lambdamin}
\lambda_{\rm min}=1+\min_{j=1,2}\left[\frac{2\epsilon_{j}}{b+(N-3)\epsilon_{j}} \right].
\end{equation}
A close inspection of the above chain of equivalent inequalities confirms that $\lambda_{\mathrm{min}}$
given by formula (\ref{lambdamin}) is indeed the minimum defined by Eq.~(\ref{lambda}).
Before going further let us notice that in the above derivation we assumed the regularity of
the matrix $\epsilon$, i.e. $\det \epsilon\neq 0$.  By continuity, the resulting formula (\ref{lambdamin}) holds also when $\det \epsilon=0$.
Without loss of any generality we can assume that $\epsilon_1 \geq \epsilon_2$.
The minimum in Eq. (\ref{lambdamin}) is then achieved for $\epsilon_2$ and $\lambda_{\mathrm{min}}=1+2\epsilon_2/[b+(N-3)\epsilon_2]$.
This completes the calculation of the minimum symplectic eigenvalue $\mu_2$ given by Eq. (\ref{mineig}). The Gaussian localizable entanglement of the fully symmetric Gaussian state with CM (\ref{gammasym}) can be determined by inserting $\mu_{2}$
into the formula (\ref{EN}) for the logarithmic negativity.

Note that in the present case of fully symmetric states we in fact did not restrict ourselves
to the local measurements and therefore we found the optimal measurement among all Gaussian measurements on modes ${\bm C}$. Assuming $\epsilon_1 \geq \epsilon_2$, the optimal measurement is a collective measurement of the global phase quadrature $p_{C,\mathrm{global}}=\frac{1}{\sqrt{N-2}}\left(\sum_{j=1}^{N-2}p_{C_j}\right)$. This measurement can be, however, implemented locally by local homodyne detections of the $p_{C_j}$ quadrature on each mode followed by summation of the obtained measurement results.

A necessary condition for localization of a non-zero amount of entanglement is that $\det \epsilon <0$. First recall that $\beta-\epsilon$ and $\beta+(N-1)\epsilon$ are valid covariance matrices, which implies that
\begin{equation}
\det(\beta-\epsilon) \geq 1, \qquad \det[\beta+(N-1)\epsilon] \geq 1.
\label{detinequality}
\end{equation}
If $\det \epsilon \geq 0$ then either $\epsilon_1\geq 0$ and $\epsilon_2\geq 0$ (i) or $\epsilon_1 \leq 0$ and $\epsilon_2 \leq 0$ (ii). In the case (i) we have using Eq.~(\ref{lambdamin}) that $\lambda_{\mathrm{min}}\geq 1$.
Taking into account the first inequality (\ref{detinequality}) we immediatelly get that $\mu_2^2 \geq 1$ and $E_{L,G}=0$. Consider now the second case (ii). From the second inequality (\ref{detinequality}) we obtain a lower bound $\epsilon_2 \geq (1-b^2)/[b(N-1)]\equiv \epsilon_{2,\mathrm{min}}$.
The formula (\ref{mineig}) can be rewritten as follows,
\begin{equation}
\mu_2^2= \frac{(b-\epsilon_2)[b+(N-1)\epsilon_2]}{b+(N-3)\epsilon_2}(b-\epsilon_1).
\label{mutwoepsilon}
\end{equation}
For a fixed $\epsilon_1$, $\mu_2^2$ is an increasing function of $\epsilon_2$ in the interval $[\epsilon_{2,\mathrm{min}},0]$. A lower bound on $\mu_2^2$  can be obtained by inserting $\epsilon_1=0$ and $\epsilon_2=\epsilon_{2,\mathrm{min}}$ into Eq. (\ref{mutwoepsilon}) which yields
$\mu_2^2 \geq (Nb^2-1)/(2b^2+N-3)$. Taking into account that $b\geq 1$ we finally arrive at $\mu_2^2\geq 1$ hence $E_{L,G}=0$.

The present derivation of Gaussian localizable entanglement for symmetric states can be simply extended to the so called bisymmetric states \cite{Serafini05} that are
invariant under the exchange of modes $A$ and $B$ as well as under the exchange of any pair out of $N-2$
modes $C_{1}, C_{2},\ldots, C_{N-2}$. Such states are described by the CM of the form
\begin{equation}
\gamma_{\mathrm{bisym}}=\left(
\begin{array}{cccccc}
\beta & \epsilon & \tau & \ldots & \ldots & \tau  \\
\epsilon & \beta & \tau & \ldots & \ldots & \tau  \\
\tau^{T} & \tau^{T} & \alpha & \xi & \ldots & \xi \\
\vdots & \vdots & \xi & \alpha & \xi & \vdots \\
\vdots & \vdots & \vdots & \xi & \ddots & \xi \\
\tau^{T} & \tau^{T} & \xi & \ldots & \xi & \alpha\\
\end{array}
\right),
\label{gammabisym}
\end{equation}
where $\beta=\mbox{diag}(b,b),\epsilon=\mbox{diag}(\epsilon_{1},\epsilon_{2}),\alpha=\mbox{diag}(\alpha,\alpha),
\xi=\mbox{diag}(\xi_{1},\xi_{2})$ and $\tau$ are $2\times2$ matrices.
By means of a balanced beam splitter $U_{AB}$ and the array of $N-3$ beam splitters
(\ref{BSarray}) we can bring the state into the block diagonal form with one $2\times2$ block
$\gamma_{B}=\beta-\epsilon$, $N-3$ blocks $2\times2$ $\gamma_{C_{2}}=\ldots=\gamma_{C_{N-2}}=\alpha-\xi$
and one $4\times 4$ block
\begin{eqnarray}
\gamma_{AC_1}=\left(
\begin{array}{cc}
\beta+\epsilon & \sqrt{2(N-2)} \tau \\
\sqrt{2(N-2)} \tau^{T} & \alpha +(N-3)\xi
\end{array}
\right). \label{gammabisymAC}
\end{eqnarray}
Now the proof of optimality goes along exactly the same lines as in the case of fully symmetric
states with the only difference that now $d_{1}$ and $d_{2}$ are eigenvalues of the (generally non-diagonal)
matrix
\begin{equation}\label{calD}
{\cal D}=\alpha+(N-3)\xi-2(N-2)\tau^{T}[\beta+\epsilon-\lambda_{\rm min}(\beta-\epsilon)]^{-1}\tau.
\end{equation}
It turns immediately out that also in the case of the considered bisymmetric $N$-mode Gaussian states
optimal measurement strategy on modes ${\bf C}$ localizing maximum entanglement between modes $A$ and
$B$ is homodyne detection.

\section{Three-mode mixed states}

We have already seen that for $N$-mode Gaussian states possessing full or specific partial
symmetry with respect to the exchange of pairs of modes maximum two-mode entanglement is
localized by homodyne detection. The same statement holds also for pure $N$-mode states
\cite{Fiurasek07}. It is therefore natural to ask whether there are also Gaussian states
for which we can localize more entanglement by projection onto a finitely squeezed state than
if we would use homodyne detection. In this section we give an affirmative answer to this
question by constructing a three-mode mixed Gaussian state for which maximum entanglement
is localized by projection of one of modes onto the coherent state.

We start with a generic three-mode mixed Gaussian state with CM $\gamma_{ABC}$ expressed via the
$2 \times 2$ submatrices with respect to the $A|B|C$ splitting as
\begin{equation}\label{gammaABC}
\gamma_{ABC}= \left(
\begin{array}{ccc}
A & D  &  E \\
D^T & B & F \\
E^T & F^T & C
\end{array}
\right).
\end{equation}

Our goal is to find optimal Gaussian measurement on mode $C$ localizing maximum entanglement
between modes $A$ and $B$. For the sake of comparison, we will also determine the performance of the optimal homodyne detection. This latter problem amounts to optimizing the relative phase between the local oscillator and signal beam in the balanced homodyne detector such that the measurement localizes maximum entanglement between $A$ and $B$. We start with a generic measurement in subsection V.A whereas the case of homodyne detection will be solved in the  subsection V.B.

\subsection{Optimal Gaussian measurement}

If we project mode $C$ onto a generic Gaussian state with CM $\gamma_{C}^{M}$ then the CM matrix $\gamma_{AB}$ specifying the resulting
Gaussian state of modes $A$ and $B$ reads \cite{Giedke02,Eisert02}
\begin{equation}
\gamma_{AB}=
\left(
\begin{array}{cc}
A & D   \\
D^T & B
\end{array}
\right)
-
\left(
\begin{array}{c}
E    \\
F
\end{array}
\right)
(C+\gamma_{C}^{M})^{-1}
\left(E^T, F^T \right).
\label{gammaABformula}
\end{equation}
Expressing this CM in the $2\times2$ blocks as
\begin{equation}
\gamma_{AB}=
\left(
\begin{array}{cc}
\cal{A} & \cal{G}   \\
{\cal{G}}^T & \cal{B}
\end{array}
\right),
\label{gammaABblocks}
\end{equation}
the symplectic invariants needed to calculate the symplectic eigenvalues given in Eq.~(\ref{symplectic})
read as
\begin{equation}\label{invariants}
\delta=\mbox{det}{\cal A}+\mbox{det}{\cal B}-2\mbox{det}{\cal G},\quad \Delta=\mbox{det}\gamma_{AB}.
\end{equation}
Assuming without loss of generality projection onto a CM of a pure state (\ref{tildegammaC})
and the submatrix $C$ of CM (\ref{gammaABC}) to be in the standard form $C=c\openone$ we get
explicitly using Eqs.~(\ref{detsum}), (\ref{gammaABformula}) and (\ref{invariants}) the invariant $\delta$ in the form
\begin{widetext}
\begin{eqnarray}\label{delta}
\delta=\mbox{det}A+\mbox{det}B-2\mbox{det}D+
\frac{(\mbox{det}E-\mbox{det}F)^2+[c+\cosh(2r)]\mbox{Tr}\chi+\sinh(2r)[\chi_{1}\sin(2\theta)+\chi_{2}\cos(2\theta)]}{1+c^2+2c\cosh(2r)},
\end{eqnarray}
\end{widetext}
where $\chi_{1}=\chi_{12}+\chi_{21}$, $\chi_{2}=\chi_{22}-\chi_{11}$ and $\chi=2E^{T}J^{T}DJF-E^{T}J^{T}AJE-F^{T}J^{T}BJF$.

The second invariant $\Delta=\mbox{det}\gamma_{AB}$ can be conveniently calculated using the formalism of Wigner
functions. The value of the Wigner function of the initial three-mode state at the origin is equal to
\begin{equation}
W_{ABC}(\bm{0})=\frac{1}{\pi^{3}\sqrt{\det\gamma_{ABC}}}.
\label{WABCorigin}
\end{equation}
After projection of mode $C$ onto the pure Gaussian state with CM $\gamma_{C}^{M}$ we get the conditional (unnormalized)
two-mode Wigner function whose value at the origin reads
\begin{equation}
\tilde{W}_{AB}(\bm{0})=\frac{1}{\pi^{3}\sqrt{\det\gamma_{ABC}}} \frac{2 \pi}{ \sqrt{\det(\Gamma_{C}+{\gamma_{C}^{M}}^{-1})}},
\label{WABtilde}
\end{equation}
 where
$\Gamma_{C}$ is the $2 \times 2$ submatrix 
of the matrix $\Gamma_{ABC}=\gamma_{ABC}^{-1}$ expressed
with respect to the $A|B|C$ splitting,
\begin{equation}\label{gammainverse}
\Gamma_{ABC}= \left(
\begin{array}{ccc}
\Gamma_{A} & \Gamma_{D}  &  \Gamma_{E} \\
\Gamma_{D}^T & \Gamma_{B} & \Gamma_{F} \\
\Gamma_{E}^T & \Gamma_{F}^T & \Gamma_{C}
\end{array}
\right).
\end{equation}
The normalization of the Wigner function $\tilde{W}_{AB}$ can be determined as a probability of projecting the state of mode $C$ characterized by covariance matrix $C$ on a pure state with CM $\gamma_{C}^{M}$, and we obtain
\begin{equation}
P_C(\bm{0})=\frac{1}{\pi \sqrt{\det C}}\frac{2\pi}{\sqrt{\det(C^{-1}+{\gamma_{C}^{M}}^{-1})}}
\label{PC}
\end{equation}
A formula similar to Eq. (\ref{WABCorigin}) holds for the normalized Wigner function of the conditionally prepared state of modes $A$ and $B$, and we have
\begin{equation}
\frac{\tilde{W}_{AB}(\bm{0})}{P_C(\bm{0})} = \frac{1}{\pi^{2}\sqrt{\det\gamma_{AB}}}.
\label{WABnormalized}
\end{equation}
Upon combining formulas (\ref{WABtilde}), (\ref{PC}) and (\ref{WABnormalized}) one finds the invariant $\Delta$ in the form
\begin{equation}\label{Delta}
\Delta=\frac{\mbox{det}(\Gamma_{C}+{\gamma_{C}^{M}}^{-1})\mbox{det}\gamma_{ABC}}{\mbox{det}(C^{-1}+{\gamma_{C}^{M}}^{-1})\mbox{det}C}.
\end{equation}
Making use of Eq.~(\ref{detsum}) finally leads to the following expression for the invariant $\Delta$:
\begin{widetext}
\begin{eqnarray}\label{Delta2}
\Delta=\frac{\mbox{det}\gamma_{ABC}\{1+\mbox{det}\Gamma_{C}+\cosh(2r)\mbox{Tr}\Gamma_{C}-
\sinh(2r)\left[g_{1}\sin(2\theta)+g_{2}\cos(2\theta)\right]\}}{1+c^2+2c\cosh(2r)},
\end{eqnarray}
\end{widetext}
where $g_1=\Gamma_{C,12}+\Gamma_{C,21}$ and $g_{2}=\Gamma_{C,22}-\Gamma_{C,11}$.
For a generic three-mode Gaussian state both the symplectic invariants (\ref{delta}) and (\ref{Delta2}) are nontrivial functions of the squeezing $r$ and phase $\theta$. Consequently, minimization of the lower symplectic
eigenvalue $\mu_2$  given by Eq. (\ref{symplectic}) amounts to the finding of the minimum of the
function $f\equiv2\mu_2^{2}=\delta-\sqrt{\delta^{2}-4\Delta}$ of two variables $r$ and $\theta$. Extremal equations
$\partial f/\partial r=0$ and $\partial f/\partial\theta=0$ for optimal $r$ and $\theta$ yield after some algebra the following pair of equations:
\begin{equation}\label{extremal}
\Delta\left(\partial_{j}\delta\right)^2-\delta\left(\partial_{j}\delta\right)\left(\partial_{j}\Delta\right)+\left(\partial_{j}\Delta\right)^{2}=0,\quad j=r,\theta,\nonumber\\
\end{equation}
where the symbol $\partial_{j}$ stands for partial derivative with respect to the variable $j$. On inserting in Eq.~(\ref{extremal}) for $\delta$ and $\Delta$ from Eqs.~(\ref{delta}) and (\ref{Delta2}) and performing the respective derivatives one gets the optimal phase $\theta$ and squeezing $r$ to be solutions of the set of two
coupled polynomial equations,
\begin{eqnarray}
\frac{y^2}{(1+x^{2})^{3}}\sum_{i=0}^{6}\sum_{j=0}^{2}(P_{\theta})_{ij}x^{i}y^{j}=0,\label{Ptheta}\\
\frac{1}{(1+x^{2})^{3}}\sum_{i=0}^{6}\sum_{j=0}^{6}(P_{r})_{ij}x^{i}y^{j}=0,\label{Pr}
\end{eqnarray}
where $x=\tan\theta$, $y=\tanh r$ and the coefficients $\left(P_{\theta}\right)_{ij}$ and $\left(P_{r}\right)_{ij}$ are
complicated functions of elements of the CM (\ref{gammaABC}) that are not specified here. The equation (\ref{Ptheta}) can be solved with respect to the variable $y$  as
\begin{equation}\label{roots}
y_{1,2}=\frac{-s_{1}\pm\sqrt{s_{1}^{2}-4s_{0}s_{2}}}{2s_{2}},
\end{equation}
where $s_{j}$, $j=0,1,2$ are sixth-order polynomials in the variable $x$. Substituting now from Eq.~(\ref{roots}) for
$y$ into equation (\ref{Pr}) we obtain after some algebra a single polynomial equation in single variable $x$. Thus the problem of finding optimal Gaussian
measurement on a single mode of a generic three-mode mixed Gaussian state localizing maximum entanglement between the
remaining two modes boils down to the finding of roots of a high-order polynomial in a single variable which can be solved efficiently on a computer. Note, that besides the roots (\ref{roots}) one also has to consider the root $y=0$ that corresponds to projection of mode $C$ onto vacuum and the boundary case $y=\pm 1$ corresponding to homodyne detection on mode $C$. These measurements can also potentially represent the optimal entanglement localization strategy.

\subsection{Optimal homodyne detection}

Homodyne detection can be obtained as a
limiting case when $r\rightarrow\infty$. In this limit Eqs.~(\ref{delta}) and (\ref{Delta2}) simplify to
\begin{eqnarray}
\delta^{(\rm hom)}&=&\mbox{det}A+\mbox{det}B-2\mbox{det}D+\frac{\chi_{22}(\theta)}{c},\label{deltahom}\\
\Delta^{(\rm hom)}&=&\frac{\mbox{det}\gamma_{ABC}\Gamma_{C,11}(\theta)}{c},\label{Deltahom}
\end{eqnarray}
where $\chi(\theta)=U^{T}(\theta)\chi U(\theta)$ and $\Gamma_{C}(\theta)=U^{T}(\theta)\Gamma_{C}U(\theta)$.
The extremal equation $\partial f/\partial\theta=0$ leads to the equation (\ref{extremal}), where $j=\theta$,
which can be further arranged into the form
\begin{equation}\label{polyhom}
\frac{\mbox{det}\gamma_{ABC}}{c^3(1+x^{2})^{2}}\sum_{i=0}^{4}h_{i}x^{i}=0,
\end{equation}
where $x=\tan\theta$ and the coefficients $h_{i}$ can be found in the Appendix. Inspection of Eq.~(\ref{polyhom})
reveals that the problem of finding optimal phase in homodyne detection on mode $C$ that maximizes entanglement
between modes $A$ and $B$ reduces to the problem of finding roots of only fourth-order polynomial which can be solved
analytically. In addition to the roots one has to consider also the limiting case $x\rightarrow\infty$ corresponding to
the phase $\theta=\pi/2$ as a possible candidate for the extremum.

\subsection{Isotropic states}

Previous results reveal that in the case of a general Gaussian measurement the maximum localizable
entanglement can be found only numerically. However, several exceptions exist such as pure three-mode
states \cite{Fiurasek07} or bisymmetric states investigated in Sec.~\ref{symmetric}, where fully analytical
solution can be derived. Besides these states,  there is yet another class of the so-called isotropic
states \cite{Botero03} for which the optimization task can be resolved analytically.
The three-mode isotropic states with CM $\gamma_{ABC}$ are Gaussian states that can be transformed by a suitable
symplectic transformation $S$ into the normal form $S\gamma_{ABC}S^{T}=\nu\openone$, where $\nu\geq1$ is the three-fold degenerate
symplectic eigenvalue. Since $\nu=1$ for pure states, the CM of an isotropic state can be expressed as
\begin{equation}\label{isotropic}
\gamma_{ABC}=\nu\gamma_{p},
\end{equation}
where $\gamma_{p}=S^{-1}\left(S^{T}\right)^{-1}$ is CM of a pure state. Making use of the relation
$(\Omega\gamma_{p})^{2}=-\openone$ \cite{Holevo_82} that is valid for pure states we get
\begin{equation}\label{gammap}
\gamma_{p}^{-1}=-\Omega\gamma_{p}\Omega.
\end{equation}
Hence, we can calculate the inverse matrix (\ref{gammainverse}) for isotropic state as
\begin{equation}
\Gamma_{ABC}=\frac{1}{\nu}\gamma_{p}^{-1}=-\frac{1}{\nu^{2}}\Omega\gamma_{ABC}\Omega,
\end{equation}
where Eqs.~(\ref{gammap}) and (\ref{isotropic}) were used. Assuming the submatrix $C$ to be again in the standard
form $C=c\openone$ and taking into account the relation $\Omega=\oplus_{i=1}^{3}J$ we finally arrive at the
submatrix $\Gamma_{C}$ for an isotropic state in the form:
\begin{equation}\label{GammaC}
\Gamma_{C}=\frac{c}{\nu^2}\openone.
\end{equation}
Consequently, for isotropic states (with $C$ brought into the standard form) the parameters $g_{1}$ and $g_{2}$ vanish
and the symplectic invariant (\ref{Delta2}) is independent of the phase $\theta$. This observation greatly simplifies
optimization with respect to the phase. Since only the invariant $\delta$ depends on $\theta$ the extremal equation
$\partial f/\partial\theta=0$ is equivalent to the equation $\partial \delta/\partial\theta=0$ which is obviously
solved by $\cos(2\theta)=\pm\chi_{2}/\|\chi\|$ and $\sin(2\theta)=\pm\chi_{1}/\|\chi\|$,
where $\|\chi\|\equiv\sqrt{\chi_{1}^{2}+\chi_{2}^{2}}$. Calculating now the second derivative
$\partial^{2} f/\partial\theta^{2}$ one finds it is nonnegative for the plus sign and therefore the lower
symplectic eigenvalue $\mu$ is minimized by
\begin{equation}\label{optphase}
\sin(2\theta)=\frac{\chi_{1}}{\|\chi\|},\quad \cos(2\theta)=\frac{\chi_{2}}{\|\chi\|}.
\end{equation}

We see that for isotropic states we can find analytically optimal phase $\theta$ of the CM $\gamma_{C}^{M}$ even without resorting to the limit of infinitely large $r$ corresponding to homodyne detection. In this limit we get in particular
\begin{equation}\label{symplectichom}
\mu_{\rm min}^{(\rm hom)}=\sqrt{\frac{\delta_{\rm min}^{(\rm hom)}-\sqrt{(\delta_{\rm min}^{(\rm hom)})^2-4\nu^{4}}}{2}},
\end{equation}
where
\begin{equation}\label{deltahommin}
\delta_{\rm min}^{(\rm hom)}=\mbox{det}A+\mbox{det}B-2\mbox{det}D+\frac{1}{2c}\left(\mbox{Tr}\chi+\|\chi\|\right),
\end{equation}
and Eqs.~(\ref{Deltahom}), (\ref{isotropic}) and (\ref{GammaC}) were used.

In the case of projection onto the state with finite squeezing $r$ we have to solve the extremal equation
$\partial f/\partial r=0$ which leads to the equation (\ref{extremal}) with $j=r$. Substituting from Eqs.~(\ref{optphase})
and (\ref{GammaC}) into Eqs.~(\ref{delta}) and (\ref{Delta2}) and inserting the obtained symplectic invariants into
the extremal equation we finally arrive at the polynomial equation for optimal squeezing in the form
\begin{eqnarray}\label{optimalr}
\sum_{i=0}^{4}q_{i}y^{i}=0,
\end{eqnarray}
where $y=\tanh r$ and the coefficients $q_{i}$ can be found in the Appendix. Thus, the problem of finding optimal Gaussian
measurement localizing maximum entanglement for a generic three-mode isotropic state requires to calculate roots of only fourth-order polynomial which can be performed analytically.

\begin{figure}[!t!]
\centerline{\includegraphics[width=0.95\linewidth]{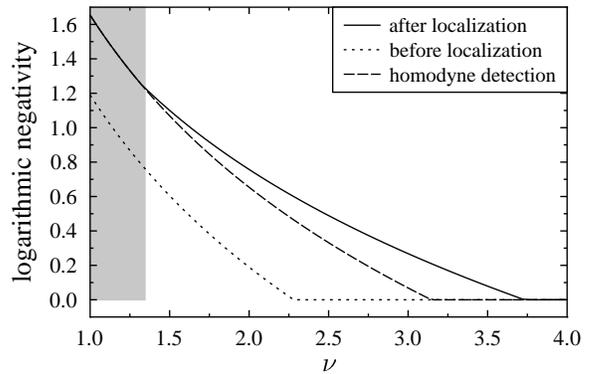}}
\caption{Dependence of entanglement of modes $A$ and $B$ on $\nu$
for the three-mode Gaussian isotropic state with generating covariance matrix (\ref{example}).
The logarithmic negativity of the two-mode state before (dotted curve) and after (solid curve) the
localization is plotted. The shaded area indicates the region where homodyne detection (dashed curve) on $C$ is the optimal
entanglement localization strategy. Otherwise, projection of mode $C$ onto coherent states is optimal.} \label{fig0}
\end{figure}

The isotropic states are of particular interest because they represent examples of states for which it is not in general optimal to perform homodyne
detection. Instead, there are isotropic states for which maximum entanglement is localized by projection onto a finitely squeezed state.
Consider the isotropic state with CM (\ref{isotropic}),
where $\nu=2$ and
\begin{equation}\label{example}
\gamma_{p}= \left(
\begin{array}{cccccc}
3 & 0  &  2 & 0 & 2 & 0 \\
0 & 3 & 0 & -2 & 0 & -2 \\
2 & 0 & 2 & 0 & 1 & 0 \\
0 & -2 & 0 & 2 & 0 & 1 \\
2 & 0 & 1 & 0 & 2 & 0 \\
0 & -2 & 0 & 1 & 0 & 2 \\
\end{array}
\right).
\end{equation}
Before measurement on mode $C$  the logarithmic negativity of the reduced bipartite state
of modes $A$ and $B$ reads $E_{\mathcal{N}}= 0.189$ e-bits.
 Further, for the CM (\ref{example}) we get $\|\chi\|=0$ and Eqs.~(\ref{symplectichom})
and (\ref{deltahommin}) give a lower eigenvalue $\mu_{2,\rm min}^{(\rm hom)}= 0.636$
corresponding to the optimal homodyne detection, yielding $E_{L,\mathrm{hom}}= 0.654$ e-bits. By analytically solving polynomial
equation (\ref{optimalr}) we find that it is optimal to take $r=0$ which corresponds to projection onto the coherent
state for which we get $\mu_{2,\rm min}= 0.592$ and $E_{L,G}= 0.757$ e-bits.
The largest entanglement is thus localized by projection onto the coherent state with CM $\gamma_{C}^{M}=\openone$ which
can be implemented by eight-port homodyne detection.
For illustration, the dependence of the localizable entanglement on the parameter $\nu$ is plotted in Fig.~\ref{fig0}.

\section{Non-Gaussian measurements}

\begin{figure}[!t!]
\centerline{\includegraphics[width=6.5cm]{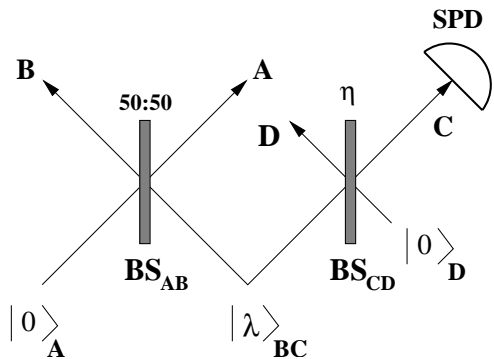}}
\caption{Scheme demonstrating superiority of non-Gaussian localization of entanglement over
optimal Gaussian localization. BS$_{AB}$: balanced beam splitter; BS$_{CD}$: unbalanced beam splitter
with transmittance $\eta$: SPD: ideal single-photon detector. See text for details.} \label{fig1}
\end{figure}

So far we have limited ourselves to Gaussian measurements. Recently, it was shown in \cite{Fiurasek07} that
this may not be optimal strategy if we allow also non-Gaussian measurements. This was illustrated on the
example of a three-mode pure Gaussian state prepared by mixing a vacuum state in mode $A$ with mode $B$ of a
two-mode squeezed vacuum state
\begin{equation}\label{lambdaBC}
|\lambda\rangle_{BC}=\sqrt{1-\lambda^2}\sum_{n=0}^{\infty}\lambda^{n}|n,n\rangle_{BC}
\end{equation}
of modes $B$ and $C$ (see Fig~\ref{fig1}), where $\lambda$ is the squeezing parameter.
Then it was demonstrated that by performing ideal photon number measurement on mode $C$ one
can localize a higher entanglement between modes $A$ and $B$ than we would get by
using optimal Gaussian measurement, i.e. homodyne detection \cite{Fiurasek07}.
While the photon number detection is suitable for the proof-of-principle of superiority of
non-Gaussian measurements it is more realistic to consider the imperfect single-photon detector
as depicted in Fig~\ref{fig1}. Here we show that this non-Gaussian measurement strategy
still outperforms (in some region of the parameter $\lambda$) the homodyne detection.

An ideal single-photon detector (SPD) has two outcomes, either a click described by the
projector $\Pi_{1}=\openone-|0\rangle\langle0|$, or no click described by the projector
onto the vacuum state $\Pi_{0}=|0\rangle\langle0|$. Imperfect SPD with detector efficiency
$\eta$ is then modeled by an unbalanced beam-splitter with transmittance $\eta$ followed
by the ideal detector (see Fig.~\ref{fig1}). No click of the detector occurs with probability $p_{0}$
and heralds preparation of modes $A$ and $B$ in the mixed state $\rho_{0}$ whereas click
is detected with probability $p_{1}=1-p_{0}$ and prepares the two modes in the state $\rho_{1}$.

\begin{figure}[!t!]
\centerline{\includegraphics[width=\linewidth]{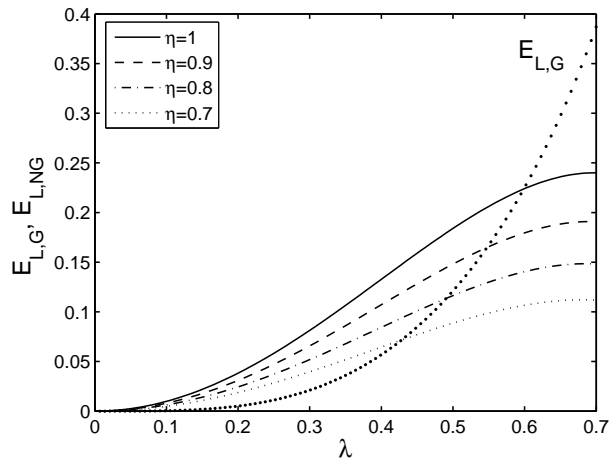}}
\caption{Gaussian localizable entanglement E$_{L,G}$ and average entanglement E$_{L,NG}$
localized by means of the single-photon detector with detector efficiency
$\eta$ are plotted versus the parameter $\lambda$. See text for details.}\label{fig2}
\end{figure}

We quantify the entanglement thus localized between modes $A$ and $B$ by the average quantity
\begin{equation}\label{average}
E_{L,NG}=p_{0}E_{\cal N}[\rho_{0}]+p_{1}E_{\cal N}[\rho_{1}].
\end{equation}
The logarithmic negativity of an arbitrary generally non-Gaussian state can be numerically calculated as $E_{\mathcal{N}}[\rho]=\log_{2}\left(1+2|\sum_{i}e_{i}|\right)$, where the sum goes over all negative eigenvalues $e_{i}$ of the partially transposed matrix $\rho^{T_{A}}$.
Projection of the mode
$C$ onto the vacuum state prepares the modes $B$ and $D$ in a two-mode squeezed vacuum state that
gives after tracing over the auxiliary mode $D$ the mode $B$ in a thermal state. Mixing of such
a state with the vacuum state $|0\rangle_{A}$ on a BS$_{AB}$ creates the state $\rho_{0}$ that
is obviously separable and hence $E_{\cal N}[\rho_{0}]=0$. It therefore suffices to calculate
the logarithmic negativity for $\rho_{1}$. Describing the beam splitters
BS$_{AB}$ and BS$_{CD}$ by the unitary operators $U_{AB}$ and $U_{CD}$, respectively,
the initial state $|\psi_{\rm in}\rangle=|0\rangle_{A}|\lambda\rangle_{BC}|0\rangle_{D}$ is transformed to
the state $|\psi\rangle_{ABCD}=U_{AB}U_{CD}|\psi_{\rm in}\rangle$. The sought conditionally prepared
(unnormalized) state then can be calculated as
\begin{equation}\label{tilderho1}
\tilde{\rho}_{1}=\mbox{Tr}_{CD}\left[|\psi\rangle_{ABCD}\langle\psi|\left(\openone_{AB}\otimes\Pi_{1,C}\otimes\openone_{D}\right)\right],
\end{equation}
and the corresponding probability reads $p_{1}=\mbox{Tr}_{AB}\left(\tilde{\rho}_{1}\right)$. After
some algebra we get explicitly the normalized density matrix $\rho_{1}=\tilde{\rho}_{1}/p_{1}$ to be
\begin{equation}\label{rho1}
\rho_{1}=\frac{\left(1-\lambda^{2}\right)}{p_{1}}\sum_{n=1}^{\infty}\lambda^{2n}\left[1-(1-\eta^2)^{n}\right]|\psi_{n}\rangle_{AB}\langle\psi_{n}|,
\end{equation}
where
\begin{equation}
|\psi_n\rangle_{AB}= \frac{1}{2^{n/2}}\sum_{k=0}^n \sqrt{n \choose k} |k,n-k\rangle_{AB},
\end{equation}
and $p_{1}=\frac{\lambda^{2}\eta^{2}}{1-\lambda^{2}(1-\eta^{2})}$. The negative eigenvalues of the
matrix $\rho_{1}^{T_{A}}$ with elements $\left(\rho_{1}^{T_{A}}\right)_{ij,kl}=\left(\rho_{1}\right)_{kj,il}$ are
calculated numerically from Eq.~(\ref{rho1}) and the obtained average logarithmic negativity
(\ref{average}) is depicted in Fig.~\ref{fig2}. For comparison we also display in the figure
maximum Gaussian localizable entanglement \cite{Fiurasek07} that is given by the entropy of entanglement
\begin{equation}
E_{L,G}=(n_A+1)\log_2(n_A+1)-n_A\log_2(n_A),
\label{entropy}
\end{equation}
where
\begin{equation}\label{nA}
n_A=\frac{1}{2}\sqrt{\frac{1}{1-\lambda^4}}-\frac{1}{2},
\end{equation}
and that is achieved by homodyne detection.
The average logarithmic negativity $E_{L,NG}$ is plotted in Fig.~\ref{fig2} as a function of the squeezing
parameter $\lambda$ for various values of detector efficiency $\eta$. We can see that even for $\eta<1$ there
is a region of squeezing parameters $\lambda$ where $E_{L,NG}> E_{L,G}$ so the imperfect SPD still
outperforms homodyne detection. The region of better performance, however, gets smaller
with decreasing detector efficiency $\eta$ and the maximum value of the difference
$E_{L,NG}-E_{L,G}$ reduces and shifts towards smaller values of $\lambda$.
The advantage of the SPD over homodyne detection can be best understood in the limit of low squeezing
$\lambda\ll1$. It follows from Eq. (\ref{nA}) that $n_{A}\approx \lambda^4/4$ and $E_{L,G}=O(\lambda^4)$.
The SPD prepares with probability $p_{1}\approx2\eta^2\lambda^2$ almost pure singlet state
$|\psi_{1}\rangle_{AB}=(|01\rangle_{AB}+|10\rangle_{AB})/\sqrt{2}$ containing one e-bit of entanglement between
$A$ and $B$ and therefore $E_{L,NG}= O(\lambda^2)$.

\section{Conclusions}

In the present paper we have analyzed in detail the localization of entanglement of mixed multimode multipartite Gaussian states by local Gaussian measurements and classical communication.
In particular, we have found that, in contrast to pure Gaussian states, the optimal Gaussian measurement that localizes maximum entanglement need not be homodyne detection (i.e. projection onto infinitely squeezed states) but rather projection on a finitely squeezed state. It is important to stress that any such measurement is experimentally feasible with present technology and can be implemented on light  by splitting the signal beam on an unbalanced beam splitter and using two balanced homodyne detectors to measure the amplitude quadrature of the first output mode and the phase quadrature of the second output mode, respectively.

The determination of the optimal Gaussian measurement for localization of entanglement of
mixed states is a non-trivial problem. Nevertheless, we were able to obtain fully analytical results for generic mixed fully symmetric Gaussian states, where we proved that homodyne detection of the amplitude or phase quadrature is optimal. For three-mode mixed Gaussian states we
have shown that the optimization problem reduces to calculation of roots of certain polynomials, which can be very efficiently performed numerically.
The problem simplifies considerably if we restrict ourselves to localization by homodyne detection where the optimal phase of the homodyne detection can be found as a root of only fourth-order polynomial which can be calculated analytically.
Fully analytical results can be obtained also for the three-mode isotropic Gaussian states
because in this case the determination of  Gaussian
localizable entanglement again  boils down to finding roots of the fourth-order polynomial.

Our results contribute to the understanding of the structure and properties of entanglement of multimode Gaussian states
and we hope that our work will stimulate the experimental investigations of entanglement localization for continuous-variable systems.

\acknowledgments
We acknowledge financial support from the Ministry of Education of the Czech Republic
(Grants No. LC06007 and No. MSM6198959213) and from GACR (Grant No. 202/08/0224).
We also acknowledge the financial support of the Future and Emerging Technologies (FET) programme
within the Seventh Framework Programme for Research of the European Commission, under the
FET-Open grant agreement COMPAS, number 212008.

\appendix*

\section{Polynomial coefficients}

The coefficients of the polynomial (\ref{polyhom}) read as
\begin{eqnarray*}
h_{4}&=&\Gamma_{C,22}\chi_{1}^2+\chi_{1}g_{1}u+cg_{1}^{2}\mbox{det}\gamma_{ABC},\\
h_{3}&=&2\chi_{1}g_{2}u+2\chi_{2}g_{1}u+4\Gamma_{C,22}\chi_{1}\chi_{2}+4cg_{1}g_{2}\mbox{det}\gamma_{ABC},\\
h_{2}&=&4\Gamma_{C,22}\chi_{2}^2-\chi_{1}^{2}\mbox{Tr}\Gamma_{C}+4\chi_{2}g_{2}u-\chi_{1}g_{1}\left(u+v\right)\\
&&+2c\left(2g_{2}^2-g_{1}^{2}\right)\mbox{det}\gamma_{ABC},\\
h_{1}&=&-4\Gamma_{C,11}\chi_{1}\chi_{2}-2v\left(\chi_{2}g_{1}+\chi_{1}g_{2}\right)-4cg_{1}g_{2}\mbox{det}\gamma_{ABC},\\
h_{0}&=&\Gamma_{C,11}\chi_{1}^{2}+\chi_{1}g_{1}v+cg_{1}^{2}\mbox{det}\gamma_{ABC},\\
\end{eqnarray*}
where
\begin{eqnarray*}
u&=&\chi_{11}+cI,\quad v=\chi_{22}+cI,
\end{eqnarray*}
and
\begin{eqnarray*}\label{I}
I=\mbox{det}A+\mbox{det}B-2\mbox{det}D.
\end{eqnarray*}

The coefficients of the polynomial (\ref{optimalr}) read
\begin{eqnarray*}
q_4&=&||\chi||^2b_{-}a_{-}, \\
q_3&=&||\chi||[2a_{-}b_{-}d_{+}- (a_{+}b_{-}+a_{-}b_{+})d_{-}], \\
q_2&=& (a_{-}b_{+}-a_{+}b_{-})^2-||\chi||^2 (a_{-}b_{+}+a_{+}b_{-})\\
& & +(d_{-}a_{+}-d_{+}a_{-})(d_{-}b_{+}-d_{+}b_{-}), \\
q_1&=& ||\chi||[2a_{+}b_{+}d_{-}- (a_{+}b_{-}+a_{-}b_{+})d_{+}], \\
q_0&=& ||\chi||^2 b_{+}a_{+},
\end{eqnarray*}
where
\begin{eqnarray*}
a_{\pm}&=& \pm \nu^2(\nu^2 \pm c)^2,\\
b_{\pm}&=& \pm (1 \pm c)^2, \\
d_{\pm}&=& \pm(1 \pm c)^2I \pm (\det E -\det F)^2 + (1\pm c) \mathrm{Tr} \chi.
\end{eqnarray*}


\end{document}